\documentclass[12pt]{article}
\usepackage{hyperref} 
\hypersetup{colorlinks=true, allcolors=black} 
\usepackage{amsfonts,amsmath,amsxtra,amsthm}
\usepackage{amssymb}
\usepackage{bm}
\usepackage[utf8]{inputenc}
\usepackage{lscape}
\usepackage[normalem]{ulem}
\usepackage[russian, english]{babel}
\usepackage{xcolor}
\usepackage{pst-circ}
\usepackage{tikz, subfig, pgfplots, ifthen}
\usepackage{tcolorbox}
\usepackage{mathrsfs}

\def\a{\alpha}

\def\b{\beta}

\def\beq{\begin{equation}}
	\def\eeq{\end{equation}}
\def\beqq{\begin{equation*}}
	\def\eeqq{\end{equation*}}

\def\bs{\begin{split}}
	\def\es{\end{split}}

\def\bu{\boldsymbol{u}}
\def\bv{\boldsymbol{v}}

\def\bk{\boldsymbol{k}}

\def\i{\hat{i}}

\def\Res{\operatorname{Res}}

\def\ve{\varepsilon}
\def\vf{\varphi}

\def\X{{\cal U}}
\def\XL{\operatorname{U}}

\def\YR{\operatorname{V}}

\def\Z{\mathbb{Z}}

	\newtheorem{lemma}{Lemma}
	
	\newtheorem{corollary}{Corollary}

	\newcommand{\rf}[1]{(\ref{#1})}

\begin{document}
		\begin{center}
			{\bf \large Hypergeometric identities related to Ruijsenaars systems}
			\bigskip
			
			{\bf N. Belousov$^{\dagger\times}$, S. Derkachov$^{\dagger\times}$, S. Kharchev$^{\bullet\ast}$, S. Khoroshkin$^{\circ\ast}$
			}\medskip\\
			$^\dagger${\it Steklov Mathematical Institute, Fontanka 27, St. Petersburg, 191023, Russia;}\smallskip\\
			$^\times${\it National Research University Higher School of Economics, Soyuza Pechatnikov 16, \\St. Petersburg, 190121, Russia;}\smallskip\\
			$^\bullet${\it National Research Center ``Kurchatov Institute'', 123182, Moscow, Russia;}\smallskip\\
			$^\circ${\it National Research University Higher School of Economics, Myasnitskaya 20, \\Moscow, 101000, Russia;}\smallskip\\
			$^\ast${\it Institute for Information Transmission Problems RAS (Kharkevich Institute), \\Bolshoy Karetny per. 19, Moscow, 127994, Russia}
		\end{center}
	\begin{abstract}
		\noindent We present a proof of	 hypergeometric identities which play a crucial role in the theory of Baxter operators in the Ruijsenaars model.
	\end{abstract}
\section{Introduction}
	\subsection{Statement}
	
	  In the paper \cite{R0} S. Ruijsenaars showed that the crucial properties of his {\it kernel function}, which serves for the solution of the Ruijsenaars system \cite{HR1,HR2}, are given by certain functional identity (now known as ``kernel function identity'') found in \cite{KN} by Y.~Kajihara  and M. Noumi. It states that for any odd function $s(z)$ of a complex variable~$z$, satisfying  the Riemann relation
	\beq\label{I0}
	\begin{split} 
		s(x+y)s(x-y)s(u+v)s(u-v)=& \, s(x+u)s(x-u)s(y+v)s(y-v)-\\ 
		& \, s(x+v)s(x-v)s(y+u)s(y-u),
	\end{split}
	\eeq
	and any complex parameter $\a$ the following identity holds
	\beq\begin{split}\label{I1}
		\sum_{\substack{I_r\subset[n] \\ |I_r|=r}}\prod_{i\in I_r}\left(\prod_{j\in [n]\setminus I_r}\frac{s(z_i-z_j-\a)}{s(z_i-z_j)}\prod_{a=1}^{n}\frac{s(z_i-y_a+\a)}{s(z_i-y_a)}\right)
		=\\ \sum_{\substack{A_r\subset[n] \\ |A_r|=r}}\prod_{a\in A_r}\left(\prod_{b\in [n]\setminus A_r}
		\frac{s(y_a-y_b+\a)}{s(y_a-y_b)}\prod_{i=1}^{n}\frac{s(z_i-y_a+\a)}{s(z_i-y_a)}\right).
	\end{split}\eeq
	Here $[n]$ denotes the set
	$$ [n]=\{1,\ldots,n\}\, .$$
	In \cite{BDKK}, studying the Baxter operators in hyperbolic Ruijsenaars system, we found that the fundamental properties of these  Baxter operator are governed by another functional identities of hypergeometric type, generalizing \rf{I1} in rational ($s(z)=z$) and trigonometric cases ($s(z)=\sin \beta z$). Following the terminology of \cite{KN}, one can name them as certain ``duality transformations for multiple hypergeometric series''.
	 In rational case they read as
	 	\beq\label{I2}\begin{split} \sum_{|\bk|=K}\prod_{i=1}^n\frac{(1+\a)_{k_i}}{k_i!}\prod_{\substack{i,j=1 \\ i\not=j}}^n\frac{(x_i-x_j-k_j-\a)_{k_i}}{(x_i-x_j-k_j)_{k_i}}\prod_{a,j=1}^n\frac{(x_j-y_a+\a)_{k_j}}
	 		{(x_j-y_a)_{k_j}}=\\
	 		\sum_{|\bk|=K}\prod_{a=1}^n\frac{(1+\a)_{k_a}}{k_a!}\prod_{\substack{a,b=1 \\ a\not=b}}^n
	 		\frac{(y_a-y_b-k_a-\a)_{k_b}}{(y_a-y_b-k_a)_{k_b}}\prod_{j,a=1}^n\frac{(x_j-y_a+\a)_{k_a}}
	 		{(x_j-y_a)_{k_a}}.
	 	\end{split}\eeq
  Here the sum is taken over all $n$-tuples
 	\beq\label{I3}\bk=(k_1,\ldots, k_n), \qquad k_i\geq0,\qquad k_1+\ldots+ k_n=K \eeq
 	of non-negative integers such that their sum equals $K$, and
 		$$(x)_n=x(x+1)\cdots (x+n-1)$$
 	is the Pochhammer symbol.
 	
 	The trigonometric version we write down with a help of $q$-analogs   $(z;q)_n$  of the Pochhammer symbol,
 	\beq\label{I4}(z;q)_n=(1-z)(1-qz)\cdots(1-q^{n-1}z).\eeq
 	Then using the same notation \rf{I3} for the summation we have
 		\beq\label{I5}\begin{split}  &\sum_{{|\bk|=K}}\prod_{i=1}^n\frac{(qt;q)_{k_i}}{(q;q)_{k_i}}	
 			\times \prod_{\substack{i,j=1 \\ i\not=j}}^n
 			\frac{(t^{-1}q^{-k_j}u_i/u_j;q)_{k_i}}{(q^{-k_j}u_i/u_j;q)_{k_i}}
 			\times
 			\prod_{a,j=1}^n\frac{(tu_j/v_a;q)_{k_j}}{(u_j/v_a;q)_{k_j}}
 			=\\
 			&\sum_{{|\bk|=K}}\prod_{a=1}^n\frac{(qt;q)_{k_a}}{(q;q)_{k_a}}	
 			\times \prod_{\substack{a,b=1 \\ a\not=b}}^n
 			\frac{(t^{-1}q^{-k_a}v_a/v_b;q)_{k_b}}{(q^{-k_a}v_a/v_b;q)_{k_b}}
 			\times
 			\prod_{a,j=1}^n\frac{(tu_j/v_a;q)_{k_a}}{(u_j/v_a;q)_{k_a}}.
 		\end{split}\eeq
 A sketch of the proof of \rf{I5} is given in \cite{BDKK}. In this note we present the complete   proof with all necessary technical details.

\subsection{Other sources and proofs}
After the first version of this note came out O. Warnaar and H. Rosengren informed us that the identity \rf{I5} in its more general elliptic form \rf{A6} has already appeared in the papers \cite[Corollary 4.3]{LSW}, \cite[eq. (6.7)]{HLNR}. The proofs in these papers are different from ours. They are derived from the original Ruijsenaars identity \cite{R1} (or from the related one) on the single tuple of variables, responsible for the commutativity of Ruijsenaars-Macdonald operators
\beq\begin{split}\label{I6}
 &	\sum_{\substack{I_r\subset[n] \\ |I_r|=r}} \, \prod_{\substack{i\in I_r\\ j\not\in  I_r}}\frac{s(x_i-x_j-\a)s(x_i-x_j+\a-\beta)}{s(x_i-x_j)s(x_i-x_j-\beta)}=\\
 &\sum_{\substack{I_r\subset[n] \\ |I_r|=r}} \, \prod_{\substack{i\in I_r\\ j\not\in  I_r}}\frac{s(-x_i+x_j-\a)s(-x_i+x_j+\a-\beta)}{s(-x_i+x_j)s(-x_i+x_j-\beta)}
\end{split}\eeq
using multiple principal specialization technique, see \cite{KN,K}. Since the proofs of the identity are quite different we  leave our note  in its original form.

 \setcounter{equation}{0}
\section{Proof} The proofs of \rf{I2} and  \rf{I5} are similar. In fact, the identity \rf{I2} can be obtained as the limit of \rf{I5}, so we prove \rf{I5}. For the proof
 	it is more convenient to rewrite the identity \rf{I5} in terms of symmetric $q$-analogs of the Pochhammer symbols,
 	\beq\label{Id1} [z;q]_n=(z^{1/2}-z^{-1/2})(q^{1/2}z^{1/2}-q^{-1/2}z^{-1/2})\cdots (q^{(n-1)/2}z^{1/2}-q^{(-n+1)/2}z^{-1/2}).
 	\eeq
 	Then \rf{I5} becomes
 	\beq\label{p4}\begin{split}  &\sum_{|\bk|=K}\prod_{i=1}^n\frac{[qt;q]_{k_i}}{[q;q]_{k_i}}	
 		\times \prod_{\substack{i,j=1\\i\not=j}}^n
 		\frac{[t^{-1}q^{-k_j}u_i/u_j;q]_{k_i}}{[q^{-k_j}u_i/u_j;q]_{k_i}}
 		\times
 		\prod_{a,j=1}^n\frac{[tu_j/v_a;q]_{k_j}}{[u_j/v_a;q]_{k_j}}
 		=\\
 		&\sum_{|\bk|=K}\prod_{a=1}^n\frac{[qt;q]_{k_a}}{[q;q]_{k_a}}	
 		\times \prod_{\substack{a,b=1 \\ a\not=b}}^n
 		\frac{[t^{-1}q^{-k_a}v_a/v_b;q]_{k_b}}{[q^{-k_a}v_a/v_b;q]_{k_b}}
 		\times
 		\prod_{a,j=1}^n\frac{[tu_j/v_a;q]_{k_a}}{[u_j/v_a;q]_{k_a}}.
 	\end{split}\eeq
 The proof uses the standard arguments from the complex analysis: we check in a rather tricky way
 that the difference between the left and right hand sides has zero residues at all possible simple poles. Thus, both sides are the polynomials symmetric over the variables $u_i$ and over the variables  $v_j$. Then the  asymptotic analysis of these polynomials shows that their difference is actually equal to zero.

 The crucial step --- calculation of the residues of both sides of the equality --- divides into two parts. First we show that each side is regular at the diagonals $u_i=q^pu_j$ and $v_a=q^sv_b$ between the variables of the same group, see Lemma \ref{lemmap1}. In this calculation we actually observe the canceling of  terms grouped in corresponding pairs. Then we show that residues at mixed diagonals $u_i=q^pv_a$ vanish. This is done by induction, using the nontrivial relation between such residues stated in Lemma \ref{lemmap2}.

 During the calculations we use the following properties of symmetric $q$-Pochhammer symbols
 \begin{align}\label{p8a}
 	&[q^pu;q]_m\times[u]_n=[q^pu]_{n-p}\times[u]_{m+p},\\[6pt]
 	\label{p8}
 	&[qu;q]_m\times [q^{-(m+p)}u^{-1};q]_n=(-1)^p[qu;q]_{m+p}\times[q^{-m} u^{-1};q]_{n-p}
 \end{align}
 which are valid for any $u$ and integer $m,n,p$.
 Here we assume that
 \beq\label{p9}
 [z;q]_{-n}=(q^{1/2}z^{1/2}-q^{-1/2}z^{-1/2})^{-1}\cdots (q^{n/2}z^{1/2}-q^{-n/2}z^{-1/2})^{-1}, \qquad  n>0.
 \eeq
 It is not difficult to verify that all the poles in \rf{p4} are simple (for generic parameter values). Consider the left hand side of \rf{p4} as the function of $u_1$ and calculate the residue of this function at the point
 \beq \label{p21} u_1=u_2q^p,\qquad p\in\Z.\eeq
 For each $\bk$, $\sum_{j=1}^nk_j=K$ denote by $\XL_{\bk}=\XL_{\bk}(\bu;\bv)$ the corresponding summand of the left hand side of \rf{p4}, and by $\YR_{\bk}=\YR_{\bk}(\bu;\bv)$ the corresponding summand of the right hand side of \rf{p4},
 \begin{align} \label{Uk}
 	&\XL_{\bk}=\prod_{i=1}^n\frac{[qt;q]_{k_i}}{[q;q]_{k_i}}	
 	\times \prod_{\substack{i,j=1 \\ i\not=j}}^n
 	\frac{[t^{-1}q^{-k_j}u_i/u_j;q]_{k_i}}{[q^{-k_j}u_i/u_j;q]_{k_i}}
 	\times
 	\prod_{a,j=1}^n\frac{[tu_j/v_a;q]_{k_j}}{[u_j/v_a;q]_{k_j}},\\
 	&\YR_{\bk}=\prod_{a=1}^n\frac{[qt;q]_{k_a}}{[q;q]_{k_a}}	
 	\times \prod_{\substack{a,b=1 \\ a\not=b}}^n
 	\frac{[t^{-1}q^{-k_a}v_a/v_b;q]_{k_b}}{[q^{-k_a}v_a/v_b;q]_{k_b}}
 	\times
 	\prod_{a,j=1}^n\frac{[tu_j/v_a;q]_{k_a}}{[u_j/v_a;q]_{k_a}}.
 \end{align}
 The summands $\XL_{\bk}$, which contribute to the residue at the point \rf{p21}, are divided into two groups.
 The denominators of the terms $\XL_{\bk}$ from the group $\bk\in I_p$ contain Pochhammer symbol
 \beqq [q^{-k_2}u_1/u_2;q]_{k_1} \eeqq
 which vanishes at the point \rf{p21}. It happens when
 \beqq
 k_2-k_1+1\leq p\leq k_2,
 \eeqq
 so that
 \beqq
 I_p=\{\bk,\, |\bk|=K: k_1\geq k_2+1-p,\ k_2\geq p\}.\eeqq
 The denominators of the terms $\XL_{\bk}$ in the group $II_p$ contain Pochhammer
 \beqq [q^{-k_1}u_2/u_1;q]_{k_2} \eeqq
 which vanishes at the point \rf{p21}. It happens when
 \beqq
 -k_1\leq p\leq k_2-k_1-1,
 \eeqq
 so that
 \beqq
 II_p=\{\bk,\, |\bk|=K: k_1\geq-p, k_2\geq k_1+1+p\}.\eeqq
 Define the maps of sets $\phi_p\colon I_p\to II_p$ and $\psi_p\colon II_p\to I_p$ by the same formulas
 \beqq
 \begin{split}
 	&\phi_p\colon I_p\to II_p \qquad \phi_p(k_1,k_2,\bk')=(k_2-p,k_1+p,\bk'),\\[6pt]
 	&\psi_p\colon II_p\to I_p \qquad \psi_p(k_1,k_2,\bk')=(k_2-p,k_1+p,\bk')\end{split}\eeqq
 where $\bk' = (k_3, \dots, k_n)$.
 \begin{lemma}\label{lemmap1}${}$
 	\begin{enumerate}\item Maps $\phi_p$ and $\psi_p$ establish bijections between the sets $I_p$ and $II_p$;
 		\item For any $\bk\in I_p$
 		\begin{align}\label{p25b} \Res_{u_1=u_2q^p}\XL_{\bk}(\bu;\bv)&+\Res_{u_1=u_2q^p}\XL_{\phi_p(\bk)}(\bu;\bv)=0,\\[6pt]
 			\Res_{v_2=v_1q^p}\YR_{\bk}(\bu;\bv)&+\Res_{v_2=v_1q^p}\YR_{\phi_p(\bk)}(\bu;\bv)=0.
 			\label{p25a}\end{align}
 	\end{enumerate}
 \end{lemma}
 \noindent {\bf Proof of Lemma \ref{lemmap1}.} The first part is purely combinatorial and can be checked directly. Let us prove the second part.

 Note first that each summand $\XL_{\bk}(\bu;\bv)$ \rf{Uk} has the following structure
 \beq\label{p10} \XL_{\bk}(\bu;\bv)=\frac{\X_{\bk}(\bu;\bv;t)}{\X_{\bk}(\bu;\bv;1)}\eeq
 where
 \beq\label{p11}
 \X_{\bk}(\bu;\bv;t)=\prod_{i=1}^n{[qt;q]_{k_i}}	
 \times \prod_{\substack{i,j=1 \\ i\not=j}}^n
 {[t^{-1}q^{-k_j}u_i/u_j;q]_{k_i}}
 \times
 \prod_{a,j=1}^n{[tu_j/v_a;q]_{k_j}}.\eeq
 We now establish the  identity
 \beq\label{p13} \X_{k_1,k_2,\bk'}(\bu;\bv;t)|_{u_1=q^pu_2}=\X_{k_2-p,k_1+p,\bk'}(\bu;\bv;t)|_{u_1=q^pu_2} = \X_{\phi_p(\bk)}(\bu;\bv;t)|_{u_1=q^pu_2}
 \eeq
 valid for any  $\bk=(k_1,k_2,\bk')\in I_p$
  with a help of an explicit bijection between linear factors of the products in both sides of the equality \rf{p13}.  All the factors in both sides of \rf{p13} which do not depend on the variables $u_1$ and $u_2$ and do not contain indices $k_1$ and $k_2$ are equal tautologically, so that the relation \rf{p13} is reduced to the  equality
  \beq \label{p14}A\prod_{j=3}^nB_j\prod_{a=1}^nC_a=A'\prod_{j=3}^nB'_j\prod_{a=1}^nC'_a
  \eeq
  where
  \beqq\begin{split}A&=[tq;q]_{k_1}\cdot [tq;q]_{k_2}\cdot[t^{-1}q^{-k_2}u_1/u_2]_{k_1}\cdot [t^{-1}q^{-k_1}u_2/u_1]_{k_2}\\
  	&=[tq;q]_{k_1}\cdot [tq;q]_{k_2}\cdot[t^{-1}q^{-k_2+p}]_{k_1}\cdot [t^{-1}q^{-k_1-p}]_{k_2};\\[6pt]
  	A'&=[tq;q]_{k_2-p}\cdot [tq;q]_{k_1+p}\cdot [t^{-1}q^{-k_1-p}u_1/u_2]_{k_2-p}\cdot [t^{-1}q^{-k_2+p}u_2/u_1]_{k_1+p}\\
  	&=[tq;q]_{k_2-p}\cdot [tq;q]_{k_1+p}\cdot [t^{-1}q^{-k_1}]_{k_2-p}\cdot [t^{-1}q^{-k_2}]_{k_1+p};
  \end{split}\eeqq
  \beqq\begin{split}B_j&=[t^{-1}q^{-k_j}u_1/u_j;q]_{k_1}\cdot [t^{-1}q^{-k_j}u_2/u_j;q]_{k_2}\cdot[t^{-1}q^{-k_1}u_j/u_1;q]_{k_j} \cdot[t^{-1}q^{-k_2}u_j/u_2;q]_{k_j}\\
  	&=[t^{-1}q^{-k_j+p}u_2/u_j;q]_{k_1}\cdot [t^{-1}q^{-k_j}u_2/u_j;q]_{k_2}\cdot	
  	[t^{-1}q^{-k_1-p}u_j/u_2;q]_{k_j}\cdot [t^{-1}q^{-k_2}u_j/u_2;q]_{k_j};\\[6pt]
  	B'_j&=[t^{-1}q^{-k_j}u_1/u_j;q]_{k_2-p}\cdot [t^{-1}q^{-k_j}u_2/u_j;q]_{k_1+p}\cdot[t^{-1}q^{-k_2+p}u_j/u_1;q]_{k_j} \cdot[t^{-1}q^{-k_1-p}u_j/u_2;q]_{k_j}\\
  	&=[t^{-1}q^{-k_j+p}u_2/u_j;q]_{k_2-p}\cdot [t^{-1}q^{-k_j}u_2/u_j;q]_{k_1+p}\cdot	
  	[t^{-1}q^{-k_2}u_j/u_2;q]_{k_j}\cdot [t^{-1}q^{-k_1-p}u_j/u_2;q]_{k_j};
  \end{split}\eeqq \vspace{4pt}
  \beqq\begin{split}C_a&=[tu_1/v_a;q]_{k_1}\cdot[tu_2/v_a;q]_{k_2}
  	=[tq^pu_1/v_a;q]_{k_1}\cdot[tu_2/v_a;q]_{k_2};\\[6pt]
  	C'_a&=[tu_1/v_a;q]_{k_2-p}\cdot[tu_2/v_a;q]_{k_1+p}
  	=[tq^pu_2/v_a;q]_{k_2-p}\cdot[tu_2/v_a;q]_{k_1+p}.
  \end{split}\eeqq
  Applications of \rf{p8a} imply the equalities
  \beqq B_j=B'_j, \qquad C_a=C'_a.\eeqq
  Applying twice \rf{p8} we get
  $ A=A'$.
  This proves \rf{p14} and, as a consequence, \rf{p13}.

   The identity \rf{p13} implies the statement \rf{p25b} about zero sum of  the residues. Indeed, the relation \rf{p13} establishes a bijection between all nonzero factors of the denominators $\X_{k_1,k_2,\bk'}(\bu;\bv;1)|_{u_1=q^pu_2}$ and $\X_{k_2-p,k_1+p,\bk'}(\bu;\bv;1)$ and the equality of their products. Factors in denominators of $\XL_{k_1,k_2,\bk'}(\bu;\bv)$ and
 $\XL_{k_2-p,k_1+p,\bk'}(\bu;\bv)$ which tend to zero when $u_1$ tends to $q^pu_2$ are
 \beq\label{p13a} q^{-p/2}u_1/u_2-q^{p/2}u_2/u_1, \qquad q^{p/2}u_2/u_1-q^{-p/2}u_1/u_2.\eeq
 They give inputs into residues, which just differ by sign. Thus, we arrive at \rf{p25b}.
 For the proof of \rf{p25a} we note that the involution
 \beq\label{p14-2} \tau\colon \; u_i\mapsto v_i^{-1}, \quad v_i\mapsto u_i^{-1}\eeq
 exchanges each $\XL_{\bk}$ with $\YR_{\bk}$.
 \hfill{$\Box$}
 \begin{corollary}\label{corollaryp1}
 	Both sides of \rf{p4} have no poles of the form $u_i=q^pu_j$ and $v_a=q^pv_b$.
 \end{corollary}

 For any non-negative integer $p$ denote  by $\vf_p(\bu;\bv)$ the following rational function
 of $\bu=(u_1,\ldots,u_n)$ and $\bv=(v_1,\ldots,v_n)$
 \beq\label{p15b}	
 \vf_p(\bu;\bv)=
 (-1)^p \frac{[tq;q]_{2p}}{[q;q]_p[q;q]_{p-1}}\prod_{j=2}^n\frac{[tu_j/v_1;q]_p}{[u_1/u_j;q]_p}\prod_{b=2}^n\frac{[tu_1/v_b;q]_p}{[v_b/v_1;q]_p}.
 \eeq
 Denote also $\bu' = (u_2,\ldots,u_n)$, and $\bv' = (v_2,\ldots,v_n)$.
 \begin{lemma}\label{lemmap2} For any $1\leq p\leq k_1$ and $\bk'\in\Z_{\geq 0}^{n-1}$
 	\begin{align}\label{p15}
 		\Res_{v_1=q^{p-1}u_1}\frac{1}{v_1}\YR_{k_1,\bk'}(\bu;\bv)=&\vf_p(\bu;\bv)\times \YR_{k_1-p,\bk'}(qv_1,\bu';q^{-1}u_1,\bv'),\\[6pt]
 		\label{p15a}
 		\Res_{v_1=q^{p-1}u_1}\frac{1}{v_1}\XL_{k_1,\bk'}(\bu;\bv)=&\vf_p(\bu;\bv)\times
 		\XL_{k_1-p,\bk'}(qv_1,\bu';q^{-1}u_1,\bv').\end{align}
 \end{lemma}
\noindent {\bf Proof of Lemma \ref{lemmap2}.} We prove \rf{p15}.  Present $\YR_{k_1,\bk'}(\bu;\bv)$ in the form
\beq\label{p16}\begin{split}
	\YR_{k_1,\bk'}&(\bu;\bv)=\frac{[tq;q]_{k_1}}{[q;q]_{k_1}}\times\\ &\prod_{b\not=1}\frac{[t^{-1}q^{-k_b}v_b/v_1;q]_{k_1}}{[q^{-k_b}v_b/v_1;q]_{k_1}}\cdot
	\frac{[t^{-1}q^{-k_1}v_1/v_b;q]_{k_b}}{[q^{-k_1}v_1/v_b;q]_{k_b}}
	\cdot\frac{[tu_1/v_b;q]_{k_b}}{[u_1/v_b;q]_{k_b}}\times\\&\prod_{j=2}^n\frac{[tu_j/v_1;q]_{k_1}}{[u_j/v_1]_{k_1}}\times \frac{[tu_1/v_1;q]_{k_1}}{[u_1/v_1;q]_{k_1}}\times \YR'
\end{split}\eeq
where $\YR'$ depends  on $\bu'$, $\bv'$, $\bk'$ only.	
Then
\beq\label{p17}\begin{split}
	&\Res_{v_1=q^{p-1}u_1}\frac{1}{v_1}\YR_{k_1,\bk'}(\bu;\bv)=C\cdot \YR'\times\\
	&\prod_{b\not=1}\frac{[t^{-1}q^{-k_b}v_b/v_1;q]_{k_1}}{[q^{-k_b}v_b/v_1;q]_{k_1}}\cdot\frac{[tu_1/v_b;q]_{p}}{[u_1/v_b]_{p}}
	\cdot\frac{[tq^pu_1/v_b;q]_{k_b-p}}{q^pu_1/v_b;q_{k_b-p}}\\
	&\prod_{b\not=1}\frac{[t^{-1}q^{-k_1}v_1/v_b;q]_{k_b}}{[q^{-k_1}v_1/v_b;q]_{k_b}}\times \prod_{j=2}^n\frac{[tu_j/v_1;q]_{p}}{[u_j/v_1;q]_{p}} \prod_{j=2}^n\frac{[tq^pu_j/v_1;q]_{k_1-p}}{[q^pu_j/v_1]_{k_1-p}}
\end{split}
\eeq
where
\beq\label{p18} C=-\frac{[tq;q]_{k_1}}{[q;q]_{k_1}}\cdot\frac{[tq^{1-p};q]_{k_1}}{[q^{1-p};q]_{p-1}
	[q^{};q]_{k_1-p}}.
\eeq
Here we decomposed two fractions of Pochhammers into the products of four fractions. In this presentation there are two products which do not depend on $k$ indices. Put them in the front and use the equality $v_1=q^{p-1}u_1$. Then the residue \rf{p17} looks as
\beq\label{p19}\begin{split}
	&\Res_{v_1=q^{p-1}u_1}\frac{1}{v_1}\YR_{k_1,\bk'}(\bu;\bv)=C\cdot \YR'\cdot\prod_{b\not=1}\frac{[tu_1/v_b;q]_{p}}{[v_1/v_b]_{p}}\cdot\prod_{j=2}^n\frac{[tu_j/v_1;q]_{p}}{[u_j/u_1;q]_{p}}\times\\
	&\prod_{b\not=1}\frac{[t^{-1}q^{-k_b}v_b/v_1;q]_{k_1}}{[q^{-k_b}v_b/v_1;q]_{k_1}}
	\cdot\frac{[tqv_1/v_b;q]_{k_b-p}}{[qv_1/v_b;q]_{k_b-p}}\times\\
	&\prod_{b\not=1}\frac{[t^{-1}q^{-k_1}v_1/v_b;q]_{k_b}}{[q^{-k_1}v_1/v_b;q]_{k_b}}\times  \prod_{j=2}^n\frac{[tqu_j/u_1;q]_{k_1-p}}{[qu_j/u_1]_{k_1-p}}.
\end{split}
\eeq
Now we use \rf{p8} in the second line of \rf{p19} together with the relation $v_1=q^{p-1}u_1$.
We get
\beq\label{p20}\begin{split}
	&\Res_{v_1=q^{p-1}u_1}\frac{1}{v_1}\YR_{k_1,\bk'}(\bu;\bv)=C\cdot \YR'\cdot\prod_{b\not=1}\frac{[tu_1/v_b;q]_{p}}{[v_b/v_1]_{p}}\cdot\prod_{j=2}^n\frac{[tu_j/v_1;q]_{p}}{[u_1/u_j;q]_{p}}\times\\
	&\prod_{b\not=1}\frac{[t^{}q^{}v_1/v_b;q]_{k_b}}{[q^{}v_1/v_b;q]_{k_b}}
	\cdot\frac{[t^{-1}q^{-k_b+p}v_b/v_1;q]_{k_1-p}}{[q^{-k_b+p}v_b/v_1;q]_{k_1-p}}\times\\
	&\prod_{b\not=1}\frac{[t^{-1}q^{-k_1}v_1/v_b;q]_{k_b}}{[q^{-k_1}v_1/v_b;q]_{k_b}}\times  \prod_{j=2}^n\frac{[tqu_j/u_1;q]_{k_1-p}}{[qu_j/u_1]_{k_1-p}}.
\end{split}
\eeq
Set
\beq v^*_1=q^{-1}u_1,\qquad u^*_1=qv_1.\eeq
Then we can read two last lines in \rf{p20} as
\beq\label{p22} \begin{split}\prod_{b\not=1}\frac{[tu^*_1/y_b;q]_{k_b}}{[u^*_1/y_b;q]_{k_b}}\cdot\frac{[t^{-1}q^{-k_1+p}v^*_1/v_b;q]_{k_b}}{[q^{-k_1+p}v^*_1/v_b;q]_{k_b}}
	\cdot	\frac{[t^{-1}q^{-k_b}v_b/v_1;q]_{k_1-p}}{[q^{-k_b}v_b/v_1;q]_{k_1-p}}\cdot\prod_{j\not=1}\frac{[tu_j/v^*_1;q]_{k_1-p}}{[u_j/v^*_1;q]_{k_1-p}}.
\end{split}\eeq
One can recognize in \rf{p22} the factor of the product $\YR_{k_1-p,\bk'}(u^*_1,\bu';v^*_1,\bv')$ with missing constant
\beq\label{p23} C'=\frac{[tq;q]_{k_1-p}}{[q;q]_{k_1-p}}\times\frac{[tu^*_1/v^*_1;q]_{k_1-p}}{[u^*_1/v^*_1;q]_{k_1-p}}=\frac{[tq;q]_{k_1-p}}{[q;q]_{k_1-p}}\times\frac{[tq^{p+1};q]_{k_1-p}}{[q^{p+1};q]_{k_1-p}}.
\eeq
We conclude that
\begin{align}\notag\label{p24}&\Res_{v_1=q^{p-1}u_1}\frac{1}{v_1}\YR_{k_1,\bk'}(\bu;\bv)=\\ \notag &\frac{C}{C'}\cdot\prod_{b\not=1}\frac{[tu_1/v_b;q]_{p}}{[v_b/v_1]_{p}}\cdot\prod_{j=2}^n\frac{[tu_j/v_1;q]_{p}}{[u_1/u_j;q]_{p}}\times \YR_{k_1-p,\bk'}(u^*_1,\bu';v^*_1,\bv')=\\
	&(-1)^p \frac{[tq;q]_{2p}}{[q;q]_p[q;q]_{p-1}}\YR_{k_1-p,\bk'}(qv_1,\bu';q^{-1}u_1,\bv').
\end{align}
The proof of \rf{p15a} is analogous. One can get it by combining the involution \rf{p14-2} with the
previous arguments. \hfill{$\Box$}
\medskip

\noindent	{\bf Proof of the identity \rf{p4}.} Now we are ready to prove \rf{p4} by induction over $K$. Denote the difference between the left and right hand sides of \rf{p4} by $W_K(\bu;\bv)$. Assume that $W_K(\bu,\bv)=0$ for all $K<N$ and any $m$-tuples of
variables $\bu= (u_1,\ldots,u_m)$, $\bv=(v_1,\ldots,v_m)$ for arbitrary $m$.
Summing up the difference of \rf{p15a} and \rf{p15} over all $\bk$ with $|\bk|=K$ we get the
relation
\beq\label{p25}\Res_{v_1=q^{p-1}u_1}\frac{1}{v_1}W_K(\bu;\bv)=\vf_p(\bu;\bv)\times W_{K-p}(\bu^*,\bv^*),
\eeq
where
\beq\label{p26}\bu^*=(qv_1,\bu'),\qquad \bv^*=(q^{-1}u_1,\bv')
\eeq
and by the induction assumption the right hand side of \rf{p25} equals zero. Taking in mind the symmetry of $W_K(\bu;\bv)$  with respect to permutation of $u_i$ and of $v_j$ we conclude that it has no poles at all.  Since $W_K(\bu;\bv)$ is a  homogeneous rational function of the variables $u_i$ and $v_j$ of total degree zero, it is equal to a constant, which could depend on $q$ and $t$. To compute this constant, we consider the behavior of this function in asymptotic zone
\beq u_1\ll u_2\ll\ldots\ll u_n\ll v_n\ll v_{n-1}\ll \ldots \ll v_1.\eeq
Here both sides of \rf{p4} tend to
\beq \sum_{{|\bk_n|=K}}\prod_{i=1}^n\frac{[qt;q]_{k_i}}{[q;q]_{k_i}}\times t^{\frac{1}{2}\big( (n-1)k_1+(n-3)k_2+\ldots+(3-n)k_{n-1}+(1-n)k_n\big)}\times t^{-\frac{nK}{2}}.\eeq
Thus, $W_K(\bu;\bv)$ tends to zero in this asymptotic zone and so equals zero identically.
Another way to verify the vanishing of  the constant value of $W_K(\bu;\bv)$ is to consider
$W_K(\bu;\bv)$ at the plane
\beq\label{p28} u_i=t^{-1}v_i,\qquad i=1,\ldots, n,\eeq
where it is identically zero due to the last products in each summands.
 This completes the induction step and the proof of the identity \rf{p4} and thus of \rf{I5}. \hfill{$\Box$}
\setcounter{equation}{0}
\section{Comments} {\bf 1}. Note first that the trigonometric kernel function identity \rf {I1} is a particular limit of the trigonometric hypergeometric identity \rf{I5}, as well as the rational kernel function identity is a particular limit of the rational hypergeometric identity \rf{I2}.

   Rescale simultaneously all the variables $x_j$, $y_a$ and $\alpha$ in \rf{I2}
 $$x_j\to \ve x_j,\qquad y_a\to \ve y_a,\qquad \a\to \ve\a,$$
 and tend the rescaling constant $\ve$ to zero. In this limit the relation \rf{I2} becomes
 \beq\label{A1}\begin{split} \sum_{|\bk|=K}\prod_{i=1}^n\prod_{\substack{i,j=1 \\ k_i\not=0,\, k_j=0}}^n\frac{(x_i-x_j-\a)}{(x_i-x_j)}\prod_{\substack{a,j=1 \\ k_j\not=0}}^n\frac{(x_j-y_a+\a)}
 	{(x_j-y_a)}-\\
 	\sum_{|\bk|=K}\prod_{\substack{a,b=1 \\ k_a=0,\, k_b\not=0}}^n
 	\frac{(y_a-y_b-\a)}{(y_a-y_b)}\prod_{\substack{j,a=1 \\ k_a\not=0}}^n\frac{(x_j-y_a+\a)}
 	{(x_j-y_a)}=0.
 \end{split}\eeq
 Denote by ${\mathcal H}_K$ the left hand side of \rf{A1} and by ${\mathcal K}_r$ the left hand side of the $r$-th rational kernel function identity:
 \beq\begin{split}\label{A2}
 	&\sum_{\substack{I_r\subset[n] \\ |I_r|=r}}\prod_{i\in I_r}\left(\prod_{j\in [n]\setminus I_r}\frac{x_i-x_j-\a}{x_i-x_j}\prod_{a=1}^{n}\frac{x_i-y_a+\a}{x_i-y_a}\right)
 	-\\ 
 	&\sum_{\substack{A_r\subset[n] \\ |A_r|=r}}\prod_{a\in A_r}\left(\prod_{b\in [n]\setminus A_r}
 	\frac{y_a-y_b+\a}{y_a-y_b}\prod_{i=1}^{n}\frac{x_i-y_a+\a}{x_i-y_a}\right)=0.
 \end{split}\eeq
 We see that ${\mathcal H}_1= {\mathcal K}_1$, that is the relation \rf{A1} for $K=1$ coincides with the relation \rf{A2} for $r=1$. Next
 \beq\label{A3}{\mathcal H}_2={\mathcal K}_2+{\mathcal K}_1\eeq
  where the first term in right hand side of \rf{A3} corresponds to partitions $\bk=(1,1,0,\ldots,0)$ and their permutations while the second to the  partitions $\bk=(2,0,\ldots,0)$. Thus we get \rf{A2} for $r=2$.
Going further we represent each ${\mathcal H}_K$ as a sum of ${\mathcal K}_K$ and of ${\mathcal K}_r$ with $r<K$ taken with some combinatorial coefficients. By induction we get all the relations
\rf{A2} from \rf{A1}.

In trigonometric case we put
\beqq u_i=e^{2\imath \beta x_i},\qquad v_a=e^{2\imath \beta y_a},\qquad t=e^{2\imath \beta \a},\qquad q=e^L\eeqq
and tend in \rf{p4} the positive constant $L$ to infinity. By the same arguments we get \rf{I1} for $s(z)=\sin\b z$.
Note that the original Ruijsenaars identity \rf{I6} on a single tuple of variables could not be derived from the identity \rf{I1} on the two tuples of variables. Probably, the same negative statement holds  for the identity \rf{I5}.

\medskip

{\bf 2}. The hypergeometric identities \rf{I5} remains valid, if we replace the $q$-Pochhammer symbol  \rf{I4} by its elliptic analog
\beq \label{A4} (z;p,q)_k=\theta(z;p)\theta(qz;p)\cdots\theta(q^{k-1}z;p)\eeq
where $|p|<1$ and
\beq\label{A5} \theta(z;p)=\prod_{n\geq0}(1-p^nz)\prod_{m>0}(1-p^mz^{-1})\eeq
is the modified theta function, so that the identity \rf{I5} takes the form
	\beq\label{A6}\begin{split}  &\sum_{{|\bk|=K}}\prod_{i=1}^n\frac{(qt;p,q)_{k_i}}{(q;p,q)_{k_i}}	
	\times \prod_{\substack{i,j=1 \\ i\not=j}}^n
	\frac{(t^{-1}q^{-k_j}u_i/u_j;p,q)_{k_i}}{(q^{-k_j}u_i/u_j;p,q)_{k_i}}
	\times
	\prod_{a,j=1}^n\frac{(tu_j/v_a;p,q)_{k_j}}{(u_j/v_a;p,q)_{k_j}}
	=\\
	&\sum_{{|\bk|=K}}\prod_{a=1}^n\frac{(qt;p,q)_{k_a}}{(q;p,q)_{k_a}}	
	\times \prod_{\substack{a,b=1 \\ a\not=b}}^n
	\frac{(t^{-1}q^{-k_a}v_a/v_b;p,q)_{k_b}}{(q^{-k_a}v_a/v_b;p,q)_{k_b}}
	\times
	\prod_{a,j=1}^n\frac{(tu_j/v_a;p,q)_{k_a}}{(u_j/v_a;p,q)_{k_a}}.
\end{split}\eeq
The difference   $W_K(\bu;\bv)$ between the left and right hand sides of \rf{A6} satisfies quasiperiodicity conditions
\beq\label{A7} \begin{split}W_K(u_1,\ldots, pu_i,\ldots,u_n;\bv)&=t^{-K}W_K(u_1,\ldots, u_i,\ldots,u_n;\bv),\\W_K(\bu;v_1,\ldots, pv_i,\ldots,v_n)&=t^{K}\ W_K(\bu;v_1,\ldots, v_i,\ldots,v_n).\end{split}
\eeq
	By using \rf{A7} the absence of singularities in $W_K(\bu;\bv)$ is checked in the same way as in the trigonometric case. Then, using \rf{A5} and the substitution \rf{p28} one can show that $W_K(\bu;\bv)$ vanishes identically.
	
Note finally that the identities \rf{I1}, \rf{I5} and \rf{A6} could have  a matrix generalization. The paper \cite{MZ} suggests such a possibility.

\section*{Acknowlegements}

We are grateful to Ole Warnaar and Hjalmar Rosengren 
for communicating to us their results.

The work of N. Belousov and S. Derkachov was supported by the Theoretical Physics and Mathematics Advancement Foundation «BASIS». The work of S. Khorshkin was supported by Russian Science Foundation project No 23-41-00150. He also thanks Prof. Maria Gorelik and the Weizmann Institute of Science for the kind hospitality during the summer of 2022, where the main part of this work was done.


\begin{thebibliography}{99}

	\bibitem[BDKK]{BDKK} N. Belousov, S. Derkachov, S. Kharchev, S. Khoroshkin, \textit{Baxter operators in Ruijsenaars hyperbolic system I. Commutativity of $Q$-operators}, arXiv:2303.06383 (2023).

	\bibitem[HLNR]{HLNR} M. Hallnäs, E. Langmann, M. Noumi, H. Rosengren, \textit{Higher order deformed elliptic Ruijsenaars operators}, Communications in Mathematical Physics \textbf{392}:2 (2022), 659--689.
	
	\bibitem[HR1]{HR1} M. Halln\"as, S. Ruijsenaars, \textit{Kernel functions and Bäcklund transformations for relativistic Calogero-Moser and Toda systems}, Journal of Mathematical Physics \textbf{53}:12 (2012), 123512.
	
	\bibitem[HR2]{HR2} M. Halln\"as, S. Ruijsenaars, \textit{Joint Eigenfunctions for the Relativistic Calogero–Moser Hamiltonians of Hyperbolic Type: I. First Steps}, International Mathematics Research Notices \textbf{2014}:16 (2014), 4400--4456.
	
	\bibitem[K]{K} Y. Kajihara, \textit{Euler transformation formula for multiple basic hypergeometric series of type $A$ and some applications}, Advances in Mathematics, \textbf{187}:1 (2004), 53--97.
	
	\bibitem[KN]{KN} Y. Kajihara, M. Noumi, \textit{Multiple elliptic hypergeometric series. An approach from the Cauchy determinant}, Indagationes Mathematicae \textbf{14}:3-4 (2003), 395--421.

	\bibitem[LSW]{LSW} R. Langer, M. J. Schlosser, S. O. Warnaar, \textit{Theta functions, elliptic hypergeometric series, and Kawanaka’s Macdonald polynomial conjecture}, SIGMA. Symmetry, Integrability and Geometry: Methods and Applications \textbf{5} (2009), 055.

	\bibitem[MZ]{MZ} M. Matushko, A. Zotov, \textit{On R-matrix identities related to elliptic anisotropic spin Ruijsenaars-Macdonald operators}, arXiv:2211.08529 (2022).
	
	\bibitem[R1]{R0} S. N. M. Ruijsenaars, \textit{Zero-eigenvalue eigenfunctions for differences of elliptic relativistic Calogero-Moser Hamiltonians}, Theoretical and mathematical physics \textbf{146}:1 (2006), 25--33.	
	
	\bibitem[R2]{R1} S. N. M. Ruijsenaars, \textit{Complete integrability of relativistic Calogero-Moser systems and elliptic function identities}, Communications in Mathematical Physics \textbf{110} (1987), 191--213.
	
\end{thebibliography}
\end{document}